\documentclass[prd,twocolumn,superscriptaddress]{revtex4}
\usepackage[utf8]{inputenc}

\usepackage{amsmath}
\usepackage{hyperref}
\usepackage{amssymb}
\usepackage{slashed}
\usepackage{float}
\usepackage{verbatim}
\usepackage{graphicx}

\begin{document}

\title{Extending theories on muon-specific interactions}

\author{Carl E. Carlson}\email{carlson@physics.wm.edu}
\affiliation {Nuclear and Particle Theory Group, College of William and Mary, Williamsburg, Virginia 23187, U.S.A.}

\author{Michael Freid}\email{mcfreid@email.wm.edu}
\affiliation {Nuclear and Particle Theory Group, College of William and Mary, Williamsburg, Virginia 23187, U.S.A.}
\affiliation{Thomas Jefferson National Accelerator Facility, Newport News, Virginia 23606}

\vskip 0.5cm
\date{August 7, 2015
}

\begin{abstract}

The proton radius puzzle, the discrepancy between the proton radius measured in muonic hydrogen and electronic hydrogen, has yet to be resolved.  There are suggestions that beyond the standard model (BSM) physics could resolve both this puzzle and the muon anomalous magnetic moment discrepancy.  Karshenboim \textit{et al.}~point out that simple, nonrenormalizable, models in this direction involving new vector bosons have serious problems when confronting high energy data.  The prime example is radiative corrections to $W\to \mu\nu$ decay which exceed experimental bounds.  We show how embedding the model in a larger and arguably renormalizable theory restores gauge invariance of the vector particle interactions and controls the high energy behavior of decay and scattering amplitudes.  Thus BSM explanations of the proton radius puzzle can still be viable. 

\end{abstract}

\maketitle

\section{Introduction}

The 7$\sigma$ discrepancy between the proton radius measured via muonic hydrogen Lamb shift~\cite{Pohl:2010zza,Antognini:1900ns}, and measured via electron scattering or electronic hydrogen atomic level splittings (including the Lamb shift in ordinary hydrogen)~\cite{Mohr:2012tt}, remains unresolved.  Assuming the discrepancy is not due to experimental error, we can consider the explanation that there exists unaccounted for physics which masks itself as a difference in proton size.  Such an effect must lower the muonic hydrogen $2S$ state by $310\ \mu$eV to match experiment.  There have been several proposed beyond-the-standard-model (BSM) theories that could explain such an effect. See~\cite{Pohl:2013yb,Carlson:2015jba} for reviews.

Furthermore, there exists a discrepancy of $3$ or more $\sigma$ between the measured muonic $g-2$ and that predicted by the standard model~\cite{Hagiwara:2011af,Jegerlehner:2011ti,Davier:2010nc,Aoyama:2012wk}.  A tantalizing hope and goal of these BSM theories is to solve both the proton radius and the muonic $g-2$ problems simultaneously.

One idea in the BSM regime is the existence of a new force that couples preferentially to muons~\cite{TuckerSmith:2010ra,Batell:2011qq,Carlson:2012pc}.  However, a number of constraints must be addressed~\cite{Barger:2010aj,Barger:2011mt,Karshenboim:2014tka}.  Many of them can be avoided if we suppose the new force carrier(s) couple only to muons on the leptonic side (or at least much more strongly to muons than to other leptons), and only to first generation particles on the hadronic side, with the hadronic coupling proportional to the electric charge.  A neat scheme of this sort has been proposed by Batell \textit{et al.}~\cite{Batell:2011qq} and further studied by Karshenboim \textit{et al.}~\cite{Karshenboim:2014tka}.  Our own study will be more generic, along the lines of models studied by Rislow and one of the present authors~\cite{Carlson:2012pc};  these models in particular have smaller electron couplings than in~\cite{Batell:2011qq,Karshenboim:2014tka} and a different pattern of parity violation such that corrections to the muonium hyperfine splitting and $^{133}\text{Cs}$ weak charge will not arise.

The new muon coupling will affect other processes involving muons, and one immediately thinks of the muon's anomalous magnetic moment, $(g-2)_\mu$, with its circa $2.1 \pm 0.7$ part-per-million (ppm) discrepancy between the standard model and experiment.  In the present context, it was early noted~\cite{TuckerSmith:2010ra} that if the mass of the BSM force carrier is very light, its effect on $(g-2)_\mu$ can explain the discrepancy directly.  If the mass of the new force carrier is not very light, one has to arrange a fine-tuning involving at least one more BSM force carrier to keep the effect of the new physics on $(g-2)_\mu$ at the correct size.  This is by now well understood~\cite{Batell:2011qq,Carlson:2012pc}, and one can even turn it into a positive by solving the $(g-2)_\mu$ problem and the proton radius puzzle with the same model.

Further, if the BSM force carrier is not too heavy, it can contribute as a bremsstrahlung or radiative correction to decays involving muons.  For example, the decay $K \rightarrow \mu\nu \phi$, where $\phi$ is the new force carrier was studied by Barger~\textit{et al.}~\cite{Barger:2011mt} for the case where $\phi$ does not visibly decay. If there is a small coupling allowing $\phi \rightarrow e^+ e^-$, then the $e^+e^-$ spectrum would show a visible bump above the QED background in $K \rightarrow \mu\nu e^+ e^-$, as  was considered in~\cite{Carlson:2013mya}. 

Karshenboim \textit{et al.}~\cite{Karshenboim:2014tka} called special attention to the need for embedding any new vector force carrier into a renormalizable theory, or at least into a theory where the scattering or decay amplitudes do not grow with energy when the energy is high.  They did this by studying radiative corrections to $W \to \mu\nu$ decay.  The width of this decay is unambiguously predicted in the standard model, and is measured with an uncertainty of about $2\%$. Karshenboim \textit{et al.}~showed that if there be just a low mass vector boson that coupled to muons (but not to $W$'s or $\nu$'s), with a coupling chosen to explain the energy splitting deficit in muonic hydrogen, then its contribution $\Gamma(W\to\mu\nu V)$ would exceed the measured width of the $W$ by a large margin.

\begin{figure}[h]
\centering
\includegraphics{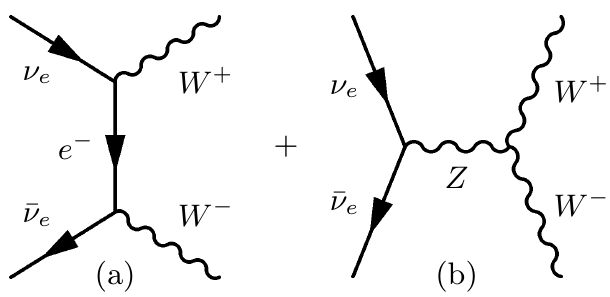}
\caption{The illustrative process $\nu \bar \nu \to W^+ W^-$.}
\label{fig:ggkl}
\end{figure}

However, in a well behaved and renormalizable theory, the growth of amplitudes with energy cannot go unchecked.
Unitarity imposes limits on the energy behavior of scattering amplitudes, and if one is using the conventions of (say) Bjorken and Drell~\cite{Bjorken:1965zz} or of Peskin and Schroeder~\cite{Peskin:1995ev}, an amplitude in a single partial wave must not grow with energy at high energy (\textit{i.e.,} if the amplitude grows like (energy)$^n$ for large energies, then $n \le 0$).  Nonrenormalizable theories are known for their ultraviolet divergences in loops, but their excessive energy dependence can also appear at tree level in the form of unitarity violations.  A known historical example is the amplitude for $\nu_e \bar\nu_e \to W^+ W^-$ in a simple vector boson theory~\cite{GellMann:1969wt}.  The calculation from just diagram~\ref{fig:ggkl}(a) gives an amplitude that is asymptotically in a single partial wave that grows like $E^2$ as the center-of-mass energy $E \to \infty$.  The Weinberg-Salam extension of the theory also has a $Z$-boson diagram,~\ref{fig:ggkl}(b), which is significantly smaller than~\ref{fig:ggkl}(a) at threshold but asymptotically cancels the offending energy behavior and restores perturbative unitarity~\cite{Weinberg:1971fb}.  A general study by Llewellyn Smith has shown that the need to satisfy unitarity bounds leads to a Yang-Mills structure for many theories involving vector bosons~\cite{Llewellyn-Smith:1973ey}.

In this paper, we consider new vector and (when needed) axial-vector bosonic interactions that couple to the muon and the proton but do not couple or couple weakly to the electron and most other particles.  Again, if this is all we have, the result of Karshenboim \textit{et al.}~shows that the region of parameter space which solves the proton radius problem does not occur in the allowed parameter space given by the known decay of the $W$.  Inspired by~\cite{Weinberg:1971fb,Llewellyn-Smith:1973ey}, we add an additional triple boson vertex in the Lagrangian,  giving an interaction involving the standard $W$-boson, the new vector particle $\phi_V$, and a further vector boson with the same mass as the $W$. We call this newest boson a ``shadow $W$,'' denoted $W_{s}$ with $m(W_{s}) = m_W$.  We also include, when needed, a corresponding axial vector triple boson interaction, involving the shadow $W_{s}$, the ordinary $W$, and the $\phi_A$.  The inclusion of the $W_s$ makes the $\phi$ interactions gauge invariant or current conserving, arguably fixes the nonrenormalizability of the original interaction, and, as we shall show, definitively pushes the constraints on the couplings due to $W$ decay far away from the coupling strength parameter region necessary to solve the proton radius problem.  Thus it can be a plausible candidate for a BSM solution to the proton radius problem.  

We note that a current conserving theory with massive bosons ($\phi_{V}$ and $\phi_{A}$) and shadow $W$'s, gives high energy results, \textit{e.g.}~for radiative corrections to $W$ decay, very much like a theory with a massive scalar boson $\phi_s$ plus, when needed, a corresponding pseudoscalar boson $\phi_p$.  We briefly display such a scalar theory, and show that decays of the $W$ involving such a scalar and pseudoscalar do not restrict the necessary parameter space needed for solving the proton radius problem with scalar exchanges.

We should also note that though our theory is well-behaved and seems likely to be renormalizable (as argued by Llewelyn Smith~\cite{Llewellyn-Smith:1973ey}), it is not yet a full theory embedded into the standard model (SM).  Further work will be required to show how such a theory can be embedded into the SM.  For now, we simply consider our theory as a phenomenological application of some BSM physics, containing features that a full theory must contain and controlling the high energy behavior of scattering and decay amplitudes.

In the following, the bulk of our work concerns the new vector or axial vector bosons, and is described in Sec.~\ref{sec:vector}.  We also include some comments on why the corresponding radiative corrections to $Z\to\mu^+\mu^-$ decay are innocuous.  Results for the scalar case are given in a short Sec.~\ref{sec:scalar}, and conclusions are offered in Sec.~\ref{sec:end}.

\section{Vector Theory}
\label{sec:vector}

We start with an interaction Lagrangian similar to~\cite{Carlson:2012pc} where $\phi_V$ interacts with a muon (and proton) via the the explicit vector coupling $C^\mu_V$ ($C^p_V$) and where $\phi_A$ interacts with a muon (and proton) through the axial vector coupling $C^\mu_A$ ($C^p_A$).  For brevity of notation, it is understood that $\phi$ without a subscript represents either $\phi_V$ or $\phi_A$ in this section.
We also include an additional 3-boson interaction~\cite{Llewellyn-Smith:1973ey} term involving the $\phi$, the ordinary $W$, and a third boson, with coupling strength equal to $C^\mu_{V}$ (or  $C^\mu_{A}$) as is necessary to make the decay $W \rightarrow \mu \nu \phi$ gauge invariant.  The third boson is the shadow $W$, denoted $W_{s}$, which couples to the muon in the same manner as the $W$ and has $m_{W_{s}} = m_W$.

The new interaction terms in the Lagrangian are,

\begin{align} \label{eq:L_int}
\mathcal{L}_{int} = &-\phi^{V}_\lambda \Big[ C^\mu_V \bar{\psi}_\mu \gamma^\lambda \psi_\mu 
	+ C^p_V \bar{\psi}_p \gamma^\lambda \psi_p  \Big]
	                \nonumber\\
		&-\phi^A_\lambda \Big[C^\mu_A \bar{\psi}_\mu \gamma^\lambda \gamma_5 \psi_\mu 
			+ C^p_A \bar{\psi}_p \gamma^\lambda \gamma_5 \psi_p \Big]
            \nonumber\\
&-i C_V^\mu \epsilon_{ijk}W^i_\alpha W^j_\beta \, \partial^\alpha W^{k,\beta}
	+i \left\{ C_A^\mu \text{\ terms} \right\}
	        \nonumber\\
	&- \frac{g}{2\sqrt{2}} 
		\bar\psi_\mu \gamma^\lambda (1-\gamma_5) \psi_\nu \, W^-_{s,\lambda}
	+ \text{h.c.}		\,,
\end{align}
where in the $C_V^\mu$ terms, 
\begin{align}
W^1_\alpha &\leftrightarrow W^-_\alpha		\,,		\nonumber\\
W^2_\alpha &\leftrightarrow W^+_{s,\alpha}	\,,		\nonumber\\
W^3_\alpha &\leftrightarrow \phi^V_\alpha		\,,		
\end{align}
with $V \to A$ for the $C_A^\mu$ terms; $\epsilon_{ijk}$ is the totally antisymmetric Levi-Civita symbol.    Note that we could use two shadow $W$'s (one vector and one axial vector), but appear to need only one.  Further note the different signs of the $C_V^\mu$ and $C_A^\mu$ Yang-Mills terms necessary for gauge invariance, and that we have included an interaction of the $W_s$ with the charge changing muon current.

If the $C^\mu_V$ and $C^p_V$ have the opposite sign then there exists an additional attractive force between the muon and the proton through the interaction with the $\phi_V$.  This additional force will create a difference between the $2S$-$2P$ Lamb shift in muonic hydrogen and hydrogen as~\cite{TuckerSmith:2010ra,Batell:2011qq,Carlson:2012pc}
\begin{align} \label{eq:lambShift}
\Delta E (2S\text{-}2P) = - \frac{ |C^\mu_V C^p_V| }{4\pi} \frac{m_\phi^2 (m_r \alpha)^3}{2(m_\phi + m_r \alpha)^4}
\end{align}
where $m_r$ is the reduced mass of the (muonic) hydrogen system.  The contribution to $\Delta E (2S\text{-}2P)$ from the axial coupling $C^\mu_A$ is very small.

\begin{figure}[h!]
\includegraphics[width=\columnwidth]{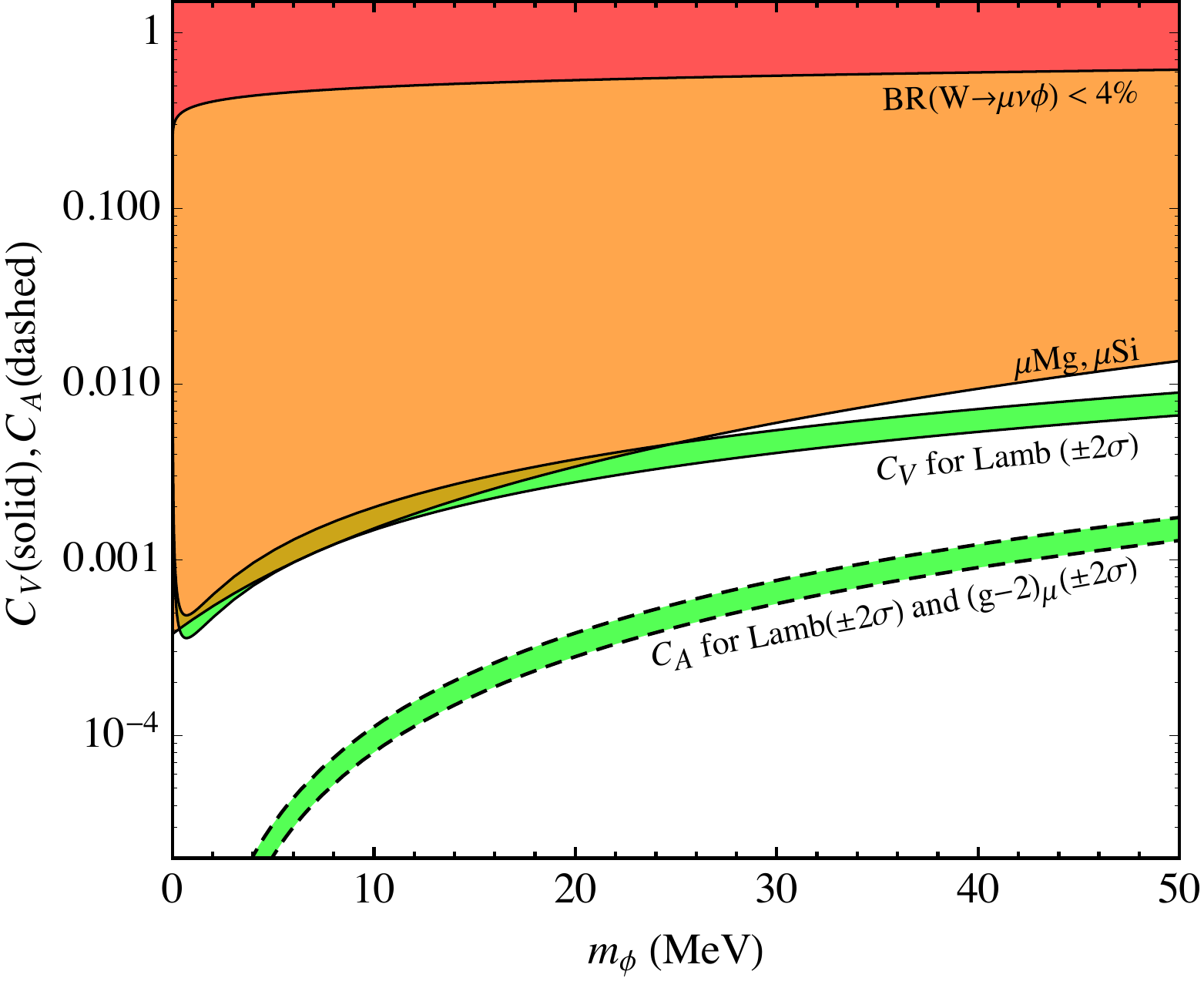}
\caption{The parameter space necessary to satisfy experimental constraints.  Solid lines refer to constraints on $C^\mu_V$.  Dashed lines refer to constraints on $C^\mu_A$.  The green band, outlined by solid lines, is the constraint on $C^\mu_V$ necessary to solve the proton radius problem ($\pm 2 \sigma$).  The shaded red region is the restricted region of $C^\mu_V$ due to the constraint that the branching ratio for $W$ goes to $\mu \nu \phi_V + \mu \nu \phi_A$ must be less than $4$\% under the assumption that $C^\mu_A$ solves the muonic $g-2$ problem.  The shaded orange region is the restricted region on $C^\mu_V$ due to energy splittings in muonic $\text{Mg}$ and $\text{Si}$ at $2 \sigma$.  The green band, outlined by dashed lines, is the constraint on $C^\mu_A$ necessary to solve the muonic $g-2$ problem ($\pm 2 \sigma$) under the assumption $C^\mu_V$ solves the proton radius problem ($\pm 2 \sigma$).  }
\label{fig:vectorConstraintsPlot}
\end{figure}

To account for the energy difference that can be interpreted as a proton radius difference, there must be an extra $310\, \mu$eV in the 2S-2P Lamb shift of muonic hydrogen~\cite{Pohl:2010zza,Antognini:1900ns}.  The parameter $C_V$ necessary to satisfy this constraint is plotted as the green band outlined by solid lines in Fig.~\ref{fig:vectorConstraintsPlot} where $|C^\mu_V| =  |C^p_V| = C_V$.

Furthermore, the introduction of new $\phi_V$ and $\phi_A$ interactions with the muon will shift the muon anomalous magnetic moment.  The vector and axial vector couplings affect the anomalous moment with opposite signs and can be tuned to account for the known discrepancy between theory and experiment of muonic $g-2$~\cite{Carlson:2012pc}.  If $C^\mu_V$ is set to satisfy the proton radius problem, then the allowed region for $C^\mu_A$ from the muon $g-2$ constraint is shown by the green band outlined by dashed lines in Fig.~\ref{fig:vectorConstraintsPlot}.

\begin{figure}[h]
\centering

\includegraphics{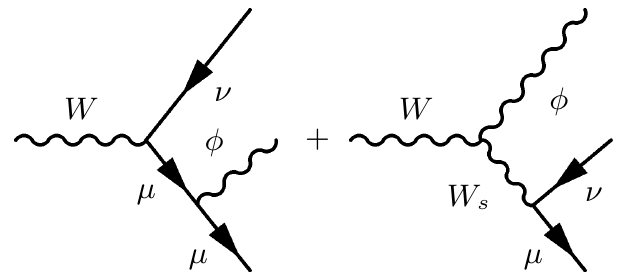}

\caption{$W \rightarrow \mu \nu \phi$}
\label{fig:WdecayVec}
\end{figure}

We now move on to consider a constraint emphasized by Karshenboim \textit{et al.}~\cite{Karshenboim:2014tka}, that the branching ratio of $W\rightarrow \mu \nu \phi_V$ plus $W\rightarrow \mu \nu \phi_A$ must be less than $4$ percent (twice the error in the $W$ width as measured by the Tevatron).  Without the inclusion of a 3-boson interaction, this constraint eliminates the region of the ($C^\mu_V$,$m_\phi$) parameter space required to explain the proton radius puzzle. This decay is calculated from the Feynman diagrams given in Fig.~\ref{fig:WdecayVec}.

From \eqref{eq:L_int} we can derive the necessary Feynman rules to compute this decay amplitude as

\begin{align}\label{eq:decayAmp}
i\mathcal{M} &= \frac{iC^\mu_V g_W}{2\sqrt{2}}\epsilon_\alpha(k)\epsilon^*_\beta(p_3)\bar{u}(p_1)\Bigg\{\frac{\gamma^\beta (\slashed{p}_1+\slashed{p}_3)}{(p_1+p_3)^2}\gamma^\alpha(1-\gamma^5)
\nonumber \\
&- \gamma^\mu(1-\gamma^5)\Bigg(\frac{g_{\mu\nu}-\frac{(p_1+p_2)_\mu (p_1+p_2)_\nu}{m_W^2}}{(p_1+p_2)^2-m_W^2}\Bigg)
\nonumber \\
&\times \Big(g^{\alpha\beta}(k+p_3)^\nu + g^{\beta\nu}(-p_3+p_1+p_2)^\alpha 
\nonumber \\
&+ g^{\alpha\nu}(-p_1-p_2-k)^\beta \Big) \Bigg\} \nu(p_2)
\end{align}
where $k$ is the $W$ 4-momentum, $p_1$ is the muon 4-momentum, $p_2$ is the neutrino 4-momentum, and $p_3$ is the $\phi_V$ 4-momentum.  Here we have focused on the vector contribution to the $W$ decay, but one can easily show that the axial vector contribution is equivalent up to an overall minus sign (which is irrelevant to the decay amplitude squared).

Letting the muon and neutrino mass be zero, we find (to leading order in $m_\phi/m_W$)
\begin{align}\label{eq:decayWidth}
\Gamma_W &= \frac{G_F m_W^3 \left[(C^\mu_V)^2+(C^\mu_A)^2\right]}{96 \sqrt{2} \pi ^3} 
                \nonumber\\
&\times \Bigg\{  \log^2 \frac{m_W^2}{m^2_\phi} - 5 \log \frac{m_W^2}{m^2_\phi}
    + \frac{37}{3} - \frac{\pi^2}{3} \Bigg\}    .
\end{align}
Keeping the muon mass would only give multiplicative corrections to the coefficients like $(1 + \mathcal O({m^2_\mu}/{m^2_W}))$. 

This decay width has a strikingly different dependence on $m_\phi$ compared to~\cite{Karshenboim:2014tka}.  The $1/m^2_\phi$ dependence found there that came from the longitudinal component of the $\phi$ polarization is canceled by the inclusion of the $W_s$ propagator in the second diagram of Fig.~\ref{fig:WdecayVec}.  Thus at leading order in $m_\phi/m_W$, the mass divergence is logarithmic and not inverse polynomial.  This logarithmic dependence pushes the constraints from $W$ decay far away from the desired parameter space of $C^\mu_V$ and $m_\phi$.

The contribution of $C^\mu_A$ in \eqref{eq:decayWidth} can be obtained in terms of $C^\mu_V$ using the constraint from $(g-2)_\mu$~\cite{Carlson:2012pc}.  The constraint from $W$ decay eliminates the region of ($C^\mu_V$,$m_\phi$) above the top curve, the shaded red area, in Fig.~\ref{fig:vectorConstraintsPlot}. The values of $C^\mu_V$  below this area are allowed by this constraint.

Another constraint on $C^\mu_V$ occurs from transitions between $3d$ and $2p$ orbitals in muonic $^{24}\text{Mg}$ and $^{28}\text{Si}$~\cite{Beltrami:1985dc,Barger:2010aj,Karshenboim:2014tka}.  At two standard deviations, this constraint is plotted as the shaded orange area bordered below by a solid black line in Fig.~\ref{fig:vectorConstraintsPlot} where allowed values of $C^\mu_V$ exist on and below this line.

Note that an additional constraint due to muonium hyperfine splitting discussed in~\cite{Karshenboim:2014tka} is not relevant here since $\phi$ does not couple to the electron (or the coupling can be kept quite small).  For similar reasons, we do not have a constraint on $C_A^\mu$ from a new parity nonconserving interaction contributing to the weak charge in $^{133}\text{Cs}$, significantly opening up the allowed parameter space for $(C_A,m_\phi)$.

In Fig.~\ref{fig:vectorConstraintsPlot} we see that there are broad regions of parameter space for which we can find values of $C^\mu_V$, $C^\mu_A$, and $m_\phi$ that simultaneously solve the proton radius puzzle and the muonic $g-2$ discrepancy while satisfying the considered experimental constraints. 

\begin{figure}[H]
\centering
\includegraphics{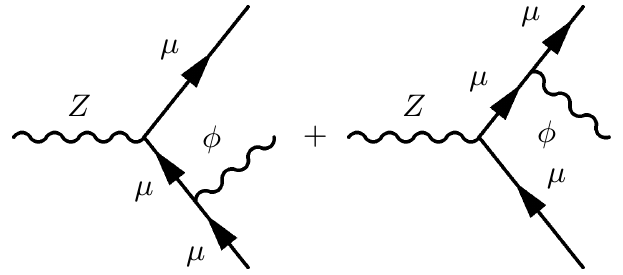}
\caption{$Z \rightarrow \mu^- \mu^+ \phi$}
\label{fig:ZdecayVec}
\end{figure}

For completeness, we also comment on radiative corrections to $Z \to \mu^+ \mu^-$ decay,  namely $Z \to \mu^- \mu^+ \phi_V$ and $\mu^- \mu^+ \phi_A$ decay as represented in Fig.~\ref{fig:ZdecayVec}.

This decay amplitude is 

\begin{align}\label{eq:decayAmpZ}
i\mathcal{M} &= \frac{i}{2} \frac{g_W}{\text{cos}\theta_W} C^\mu_V \epsilon_\alpha(k) \epsilon_\beta^*(p_3) \bar{u}(p_1)
\nonumber \\
&\times \Bigg\{ \frac{\gamma^\beta (\slashed{p}_1+\slashed{p}_3)}{(p_1+p_3)^2}\gamma^\alpha \Big(-\frac{1}{2} + 2\text{sin}^2\theta_W - \frac{1}{2}\gamma^5 \Big)
\nonumber \\
&- \gamma^\alpha \Big( -\frac{1}{2}+2\text{sin}^2\theta_W - \frac{1}{2}\gamma^5 \Big) \frac{(\slashed{p}_2 + \slashed{p}_3)\gamma^\beta}{(p_2+p_3)^2} \Bigg\} \nu(p_2)
\end{align}
where $k$ is the $Z$ 4-momentum, $p_1$ is the muon 4-momentum, $p_2$ is the anti-muon 4-momentum, and $p_3$ is the $\phi_V$ 4-momentum.  As with the $W$ decay, here we only focus on the vector contribution to the $Z$ decay, but one can easily show that the axial vector contribution is equivalent up to an overall minus sign (which is irrelevant to the decay amplitude squared).

In this case, cancellations between the two diagrams ensure the Ward identity is satisfied.  Therefore, there is no poor behavior at high energies when the $\phi$ is longitudinally polarized.  This is seen in the logarithmic dependence of the decay width \eqref{eq:decayWidthZ} on $m_\phi$ (which resembles that of the $W$ decay \eqref{eq:decayWidth}),
\begin{align}\label{eq:decayWidthZ}
\Gamma_Z &= \frac{G_F m^3_Z \left[(C^\mu_V)^2+(C^\mu_A)^2\right] \left[\frac{1}{2}-2\text{sin}^2(\theta_W)\text{cos}(2\theta_W)\right]}{48\sqrt{2}\pi^3}
\nonumber \\
&\times \Bigg\{\log^2 \frac{m^2_Z}{m^2_\phi} - 4 \log \frac{m^2_Z}{m^2_\phi} + 5 - \frac{\pi^2}{3} \Bigg\}
\end{align}

As in the calculation of the $W \rightarrow \mu \nu \phi$ decay, we have neglected the muon mass, and we have expanded the Z's decay width in \eqref{eq:decayWidthZ} to leading order in  $m_\phi/m_Z$.  These steps are motivated by the arguments given in the paragraph following \eqref{eq:decayWidth}.

\section{Scalar Theory}
\label{sec:scalar}

We also consider a scalar theory which is well behaved without the addition of any shadow particles.  The interaction Lagrangian is

\begin{align} \label{eq:L_int,S}
\mathcal{L}_{int,S} &= \phi_S \Big[ C^\mu_S \bar{\psi}_\mu \psi_\mu + C^p_S \bar{\psi}_p \psi_p \Big] \\ \nonumber
& + \phi_P \Big[ C^\mu_P \bar{\psi}_\mu \gamma^5 \psi_\mu + C^p_P \bar{\psi}_p \gamma^5 \psi_p \Big]
\end{align}
where $\phi_S$ is the scalar field, $\phi_P$ is the pseudo-scalar field where $m_{\phi_S} \equiv m_{\phi_P}$, and the $C$'s (with corresponding superscripts and subscripts) are the corresponding coupling strengths.  In this section it is understood that $\phi$ refers to either $\phi_S$ or $\phi_P$.

\begin{figure}[t]
\centering

\includegraphics{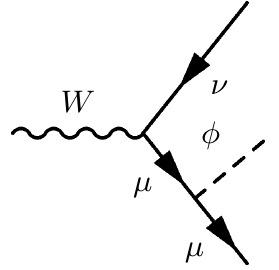}

\caption{$W \rightarrow \mu \nu \phi$}
\label{fig:WdecayScalar}
\end{figure}

As with the vector theory, we again consider the constraint due to the branching ratio of $W \rightarrow \mu \nu \phi_S$ plus $W \rightarrow \mu \nu \phi_P$.  The decay amplitude for both scalar and psuedoscalar cases is given by the Feynman diagram in Fig.~\ref{fig:WdecayScalar}.

The result for the decay width is
\begin{align}\label{eq:decayWidthScalar}
\Gamma_{W,\text{scalar}} = \frac{G_F m_W^3 \Big[ (C^\mu_S)^2 + (C^\mu_P)^2 \Big]}{96\sqrt{2}\pi^3} &\left\{ \log \left( \frac{m_W}{m_\phi} \right)
        - \frac{17}{12} \right\}
\end{align}
where, as with the vector case, we have taken the muon and neutrino mass to be zero.

\begin{figure}[t]
\includegraphics[width=\columnwidth]{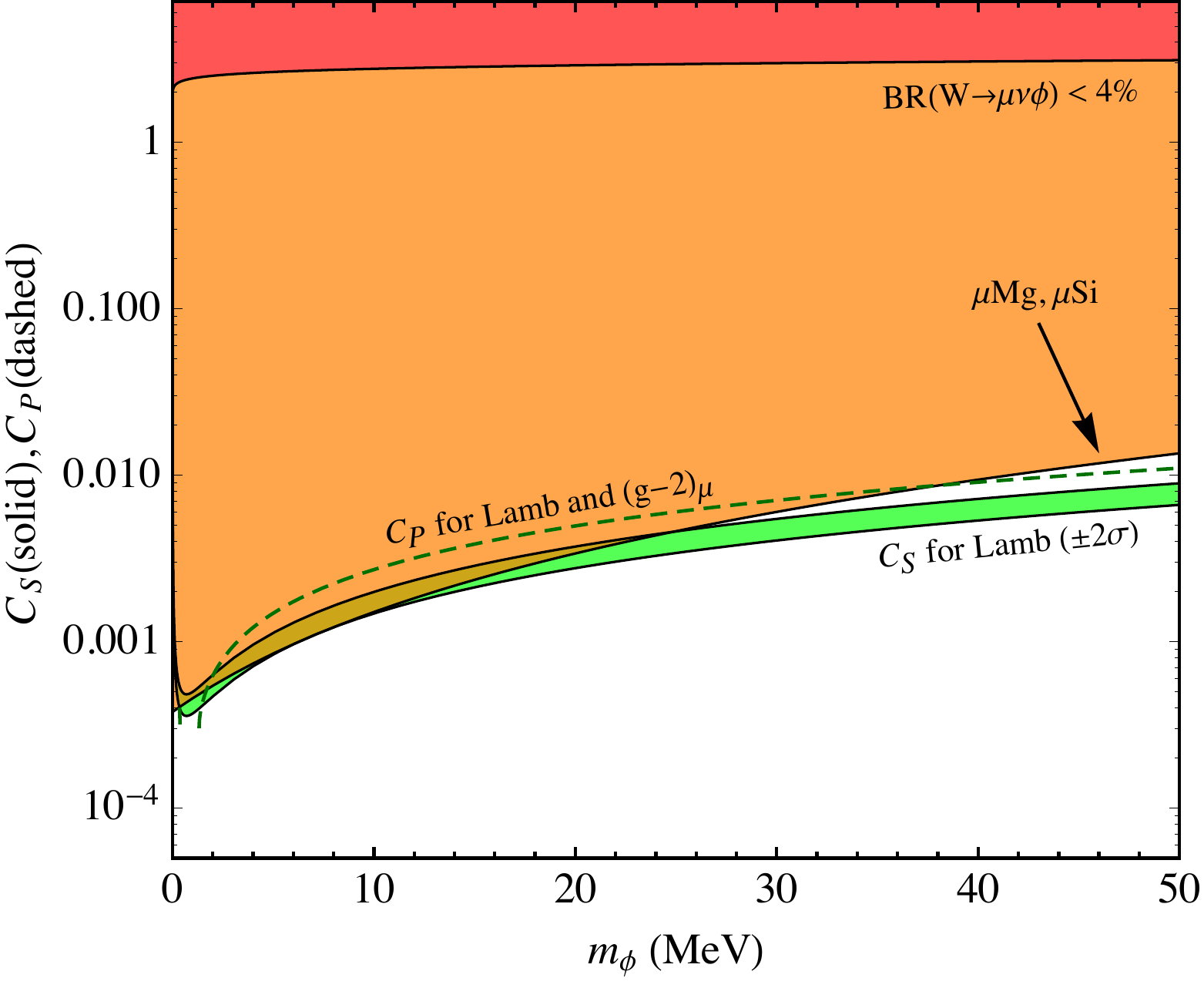}
\caption{The parameter space necessary to satisfy experimental constraints on the scalar coupling $C^\mu_S$.  The green band, outlined by solid lines, is the constraint on $C^\mu_S$ necessary to solve the proton radius problem ($\pm 2 \sigma$).  The shaded red region is the restricted region on $C^\mu_S$ due to the requirement that the branching ratio for $W$ goes to $\mu \nu \phi_S + \mu \nu \phi_P$ must be less than $4$\% under the assumption that $C^\mu_P$ solves the muonic $g-2$ problem.  The shaded orange region is the restricted region on $C^\mu_S$ due to energy splittings in muonic $\text{Mg}$ and $\text{Si}$ at $2 \sigma$.  The dashed green line is the constraint on $C^\mu_P$ necessary to solve both the proton radius and muonic $g-2$ discrepancies (for clarity, only the central value is shown).}
\label{fig:scalarConstraintsPlot}
\end{figure}

As with the vector theory, this logarithmic dependence in \eqref{eq:decayWidthScalar} opens the parameter range in $C^\mu_S$ necessary to solve the proton radius problem.  This is seen in the constraint plot Fig.~\ref{fig:scalarConstraintsPlot}.

\section{Conclusion}
\label{sec:end}

We have considered some of the constraints faced by proton radius puzzle explanations that hypothesize a new muon-specific force.

In particular, we have considered some of the necessary features required for a vector mediator to be embedded into a larger renormalizable model.  Without the embedding, amplitudes grow quickly with energy, and the energy in $W$ decay is enough to cause trouble.  The radiative corrections to $W\to \mu\nu$ would be larger than the uncertainties in the measured and well calculated $W$ decay rate, using couplings that could explain the proton radius puzzle.  However, with a theory which also contains Yang-Mills couplings, the unphysical energy dependence is ameliorated and the $W$ width constraint causes no problem for BSM models solving the proton radius puzzle.

The tightest remaining constraint follows from some measured energy splittings in muonic $\text{Mg}$ and $\text{Si}$, which are well calculated in the standard model, and new physics cannot cause a deviation larger than the experimental uncertainty.  Though this constraint is tight, it does eliminate the muon-specific interaction at the level required for the proton radius problem. 

Thus, it is fair to say that a theory involving a new vector (and axial vector) boson, $\phi$, along with vector (and axial vector) couplings to the muon and proton, as described by the Lagrangian in Eq.~(\ref{eq:L_int}), gives a viable representation of the results for some to-be-determined complete solution to the proton radius puzzle.  Such a theory necessitates the existence of an additional $W$-like boson, the shadow $W$ denoted $W_s$, which fixes the otherwise poor high energy behavior in radiative corrections to $W \rightarrow \mu \nu$ decay first noticed by Karshenboim \textit{et al.}~\cite{Karshenboim:2014tka}.

The arguable success of the vector and axial vector theory, or the scalar and pseudoscalar analog, motivates finding a suitable complete embedding into the Standard Model.  Such an embedding must be renormablizable.  The addition of a shadow $W$ and corresponding triple boson interactions is one step in this direction (as pointed out by Llewellyn Smith~\cite{Llewellyn-Smith:1973ey}).  

Finding such an embedding may necessitate new interactions which may lead to additional constraints.  For example, one must consider the process by which the new bosons acquire their mass. We also expect loop corrections to the self energy of the $W$.  Corrections with loops of new particles correlate with contributions to the Peskin-Takeuchi parameters~\cite{Peskin:1990zt,Peskin:1991sw,Hewett:1997zp}.  Though these parameters are only defined under certain conditions, \textit{e.g.,}~for BSM theories in which new particles are heavy, one can find analogous parameters for theories in which, as in our case, the new boson is light.  Contributions to these parameters will be calculated and compared against experimental constraints.  

It is also relevant to note that the addition of a new muonic interaction necessitates new radiative corrections beyond just $W$ decay.  For example, one may also consider muon decay, pion decay, and the like.  Such processes are at a significantly lighter mass scale compared to that of $W$ decay and thus are less likely to cause trouble.  We plan to asses if this is true in the near future.  Furthermore, changes to the rate of muon pair production from proton collider experiments may provide further constraints on the theory.

These are just some examples of how a vector boson theory may be more restricted than Fig.~\ref{fig:vectorConstraintsPlot} appears.  However, successes so far point to the still possible viability of these theories as solutions to the proton radius (and muonic $g-2$) puzzle.  Thus, continuation of research down this track is well motivated.

\begin{acknowledgments}

We thank Josh Erlich, Henry Lamm, Marc Sher, and Marc Vanderhaeghen for helpful conversations.  We thank the National Science Foundation for support under Grant PHY-1205905, and MF thanks the U.S. Department of Energy for support under contract DE-AC05-06OR23177, under which Jefferson Science Associates manages Jefferson Lab.

\end{acknowledgments}

\bibliography{shadow}

\end{document}